\documentstyle[floats,prb,aps]{revtex}

\begin{document}
\input epsf

\def\nd{^{\vphantom{\dagger}}} \def\ns{^{\vphantom{*}}}
\def\yd{^\dagger} \def\undertext#1{$\underline{\hbox{#1}}$}
\def\ie{{\it i.e.\/}} \def\etal{{\it et al.\/}} \def\eg{{\it
    e.g.\/}} \def\vth{\vartheta} \def\half{\frac{1}{2}}
\def\fourth{\frac{1}{4}} \def\Tr{\mathop{\rm Tr}}
\def\ket#1{{\,|\,#1\,\rangle\,}} \def\bra#1{{\,\langle\,#1\,|\,}}
\def\braket#1#2{{\,\langle\,#1\,|\,#2\,\rangle\,}}
\def\expect#1#2#3{{\,\langle\,#1\,|\,#2\,|\,#3\,\rangle\,}}
\gdef\journal#1, #2, #3, 1#4#5#6{{\sl #1~}{\bf #2}, #3 (1#4#5#6)}
\def\pra{\journal Phys. Rev. A, } \def\prb{\journal Phys. Rev. B,
  } \def\prl{\journal Phys. Rev. Lett., } \def\Nphi{{N_\phi}}
\def\abar{{\bar a}} \def\bbar{{\bar b}} \def\cS{{\cal S}}
\def\cN{{\cal N}} \def\bfr{{\mib r}} \def\bfk{{\mib k}}
\def\bfq{{\mib q}} \def\bfp{{\mib p}} \def\qhat{{\hat q}}
\def\zhat{{\hat {\bf z}}} \def\rhobar{{\overline{\rho}}}
\def\sigm{U$(N)/[$U$(m)\times$U$(N-m)$] } \def\rmL{{\rm L}}
\def\rmR{{\rm R}} \def\uar{\uparrow} \def\dar{\downarrow}
\def\klu{\ket{\rmL\uar}} \def\kld{\ket{\rmL\dar}}
\def\kru{\ket{\rmR\uar}} \def\krd{\ket{\rmR\dar}}
\def\eps{\epsilon} \def\cH{{\cal H}}

\def\IV{$I$--$V$}

\twocolumn[\hsize\textwidth\columnwidth\hsize\csname@twocolumnfalse\endcsname

\title{ Scaling Limits for the 2D Metal-Insulator Transition at
  $B = 0$ in Si-MOSFETs }

\author{ D. Lillieh\"o\"ok$^1$ and J.~E.  Furneaux$^2$ }

\address{ $^1$Department of Physics, Stockholm University, Box
  6730, S-11385 Stockholm, Sweden }

\address{ $^2$Laboratory for Electronic Properties of Matter and
  Department of Physics and Astronomy, University of Oklahoma,
  Norman, OK 73019-6021, USA }

\date{\today} \maketitle

\begin{abstract}
  We have reexamined data on the possible two dimensional
  metal-insulator transition at $B=0$ in Si-MOSFETs using a
  nonlinear regression method to extract all scaling parameters
  in a single fit. By keeping track of the magnitude of errors in
  the data we can use the normalized mean square deviation
  $\chi^2$ of the fit as a quantitative measure of how well the
  data is compatible with scaling. We have used this method to
  study electric field scaling in three different samples. We
  find rather good agreement of the data with scaling in
  individual fits, i.e. $\chi^2$'s of about 1, but also rather
  large variations in the fits depending on how cut-offs are
  introduced in the data. In particular, we report how fitted
  parameters vary when we cut away data that are either far from
  the critical point or at low excitation power, where
  temperature effects presumably dominate. In this way we find
  the critical $E$-field exponent $\beta$ to vary from about 3.2
  to 3.9 with considerably smaller statistical error estimates in
  each fit.
\end{abstract}


\vskip 1pc ] 
\narrowtext

\section{Introduction}
Several years ago the observation of a conductivity phase
transition in the two-dimensional electron system (2DES) in Si
metal-oxide-semiconductor field-effect-transistors (MOSFETs) was
reported\cite{kravchenko94d}.  This observation was unexpected
and controversial because it was contrary to the widely accepted
scaling theory for a noninteracting 2DES\cite{abrahams79}.
Further experimental studies on
MOSFETs\cite{kravchenko95b,kravchenko96a,popovic97,heemskerk98},
SiGe heterostructures\cite{lam97,coleridge97}, and GaAs
heterostructures\cite{simmons98,papadakis98,ribeiro99,hanein98b,hanein98a}
indicate that a conductivity transition is possible as a function
of carrier density in a 2DES at $B=0$.  The observed conductivity
transitions at $B=0$ bear a marked similarity to transitions
between states with different Hall resistances in the quantum
Hall effect (QHE)\cite{kravchenko95b,QHE,QUANTLIQ,sondhi97}. This
unexpected behavior and similarity to QHE transitions has also
sparked considerable theoretical
interest\cite{dob97,he98,castellani98,phillips98,belitz98,si98,altshuler99,kravchenko99,altshuler99b,klapwijk99,itp98}.
Furthermore, these parallels make it advisable to consider
developments in the QHE when studying the $B=0$ conductivity
transition in 2DES.  In particular, a recent publication by
Shahar \etal\cite{shahar98}reports that the behavior of the
resistivity as a function of temperature and electron density,
$\rho(T,n_s)$, is inconsistent with scaling at temperatures above
0.5~K.  Thus, they question the use of scaling analysis for the
study of the QHE in general.  We feel that a serious
consideration of the appropriateness of the scaling analysis for
the $B=0$ conductivity transition in 2DES is warranted.  We have
data available that can address this question using a set of data
manipulation routines in {\sl Mathematica$^{\sc tm}$} especially
developed for this purpose.  Below we briefly describe the data
used in this study and the data manipulation routines that allow
us to consistently propagate errors and eventually place
confidence limits on the applicability of this scaling analysis
to the case of the $B=0$ conductivity transition in 2DES. We feel
that these methods have applicability beyond the particular
system to the QHE and other quantum phase
transitions\cite{sondhi97}.

Our aim has been to develop an objective method that gives a
quantitative answer to the question of how well the data actually
scales. We achieve this by analyzing our ability to fit the data
to a general scaling model that assumes as little as possible
about the functional form of the scaling function. We use the
variance-weighted mean-square deviation from this fit, $\chi^2$,
as a quantitative measure of the 'goodness' of the fit. Such a
properly weighted and normalized $\chi^2$ close to one indicates
good agreement of the data with the model. An objective estimate
of $\chi^2$ requires a quantitative knowledge of experimental
uncertainties. Therefore, great care has been taken to estimate
the magnitude of these sources of uncertainty and to propagate
them consistently to the final fit.

Using this method to investigate $E$-field scaling of the lowest
temperature data available, we find rather good agreement with
scaling in the best sample with a $\chi^2$ varying from 0.4 to
1.9 depending on our selection of the data. We find that $\chi^2$
gets smaller as more data at low power is discarded. Excluding
more data at lower powers also tends to increase the consistency
among the different samples.

\section{Description of the data including sources of error}

The data used in this study was amassed at the University of
Oklahoma while studying the temperature $T$, electron density
$n_s$, and excitation dependence of the resistivity $\rho$ of a
number of high-mobility $(\mu_{\rm max} \geq 25`000~{\rm
  cm^2/Vs})$ metal-oxide-semiconductor field-effect transistors
(MOSFETs).  These devices are described in detail
elsewhere.\cite{kravchenko94d,kravchenko95b,mason96phd}

This data includes large sequences of currents $I$ with their
corresponding probe voltages $V$ recorded as a function of gate
voltage $V_g \propto n_s$ and $T$ using a four-terminal
technique. The resulting \IV curves at low temperatures are
highly nonlinear with distinctly different behaviors above and
below a critical density $n_c$ or resistivity $\rho_c$. In this
article we consider the nonlinear \IV curves taken at the lowest
temperature ($\sim 0.2$~K) and present a scaling analysis of the
resistivity as a function of the applied electric field, $E$.
This $E$ is taken to be the measured $V$ divided by the distance
between the respective probes.

\begin{figure}[htb]
  \epsfxsize 9cm \centerline{\epsffile{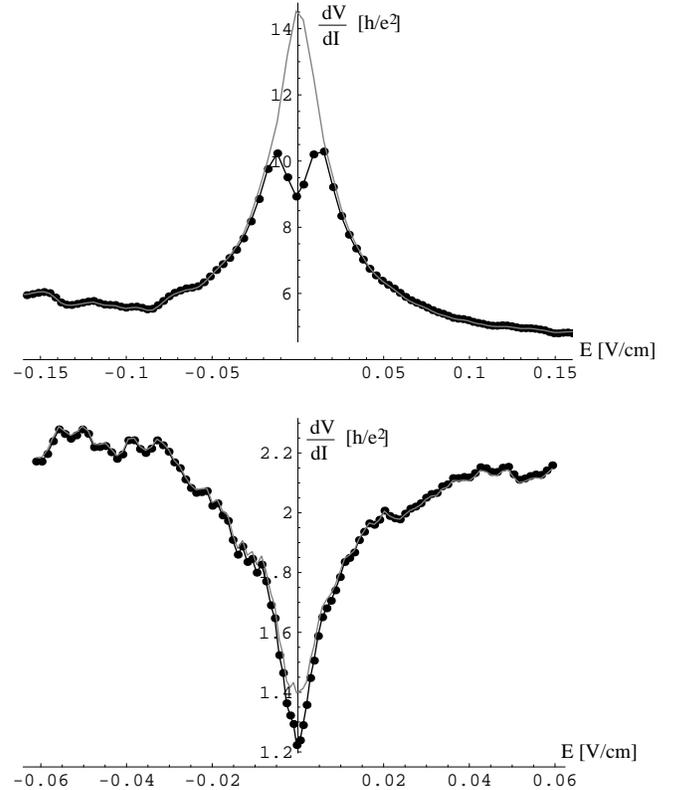}}
  \caption{Differential resistivities, $\frac{dV}{dI}$ versus
    excitation field, $E$. The upper curve taken at electron
    density $n_s=0.961*10^{11}$ cm$^{-2}$ is in the insulating
    region, whereas the lower curve at $n_s=1.05*10^{11}$
    cm$^{-2}$ shows metallic behavior. The grey curves are
    corrected for the contact anomaly by adding a small fraction
    of the contact potential drop to the measured signal.}
  \label{corrections}
\end{figure}

We find three sources of randomness that we assume are
independent:
\begin{itemize}
\item{The uncertainty in the measured probe voltage characterized
    by a root mean square (RMS) uncertainty $\sigma_V$.}
\item{The uncertainty in the measured excitation current,
    $\sigma_I$, which we find is negligible.}
\item{The uncertainty in $n_s$ due to the applied $V_g$ similarly
    characterized by $\sigma_n$.}
\end{itemize}
For these samples at this $T$ we also have a source of systematic
error that we associate with the well-known\cite{kruithof91}
difficulty in producing ohmic contacts to a 2DES.  The non-ohmic
contacts caused the measured \IV curves to be asymmetric with
respect to $I=0$ and to show a sharp anomaly centered at $I=0$.
By carefully accounting for all the voltage drops in the
measurement circuit, we were able to determine the contribution
due to the contacts.  We found that a percentage of this contact
potential (proportional to the resistivity of the sample) added
to the measured $V$ essentially eliminated the observable effects
of the contacts, see Fig. \ref{corrections}.
Furthermore, we found that these contact potentials did not
effect our analysis because they occurred at very low excitation
currents where the physics of the system is still dominated by
the finite $T$.  The results presented here are independent of
these contact anomalies and our treatment of them.

In order to measure the effective $\sigma_V$ for our data, we
produced a modified data set that was equally spaced in
excitation $I$ by using a spline interpolation scheme.  The
Fourier transform of the data was then examined. The transformed
data is consistent with a low frequency signal superimposed on
uniform white noise of constant amplitude. This white noise
amplitude varies from about 2~$\mu$V in the most metallic curves
up to 90$\mu$V in the most insulating ones. We used this
amplitude as a measure of $\sigma_V$ for each particular $V_g$ in
the following analysis.

The uncertainty in the electron density, $\sigma_n$, is more
difficult to estimate. To calibrate $n_s$ to $V_g$, we sweept the
gate voltage at a constant magnetic field to obtain
$\rho_{xx}(V_g)$.  Minima in $\rho_{xx}$ that correspond to known
integral filling factors, and therefore known $n_s$, were then
determined.  The linear fit to the resulting $(n_s, V_g)$ pairs
that gives $V_g(n_s)$ with a statistical error of $\sigma_n =
1*10^9$~cm$^{-2}$. This corresponds to a resolution of 0.8
millivolt in $V_g$.  This determination is not obviously a
measure of the density fluctuations for constant $V_g$.
Furthermore, an effective gradient in the gate voltage will be
imposed on the 2DES by the source-drain excitation.  We made
every attempt to make this potential drop symmetric with respect
to the $V$ probes, but the $n_s$ gradient could not be fully
eliminated.  Thus, we estimate the density uncertainty,
$\sigma_n$, by taking $\sigma_{V_g} \sim 1{\mbox mV}$. This
estimate is compatible with our observations of the precision and
reproducibility of producing $n_s$ with $V_g$ in our samples as
evidenced by a repeatability of a given $\rho_{xx}(V_g)$ where
$\rho_{xx}(n_s)$ is particularly steep.

\begin{figure}[hbtp]
  \epsfxsize 9cm \centerline{\epsffile{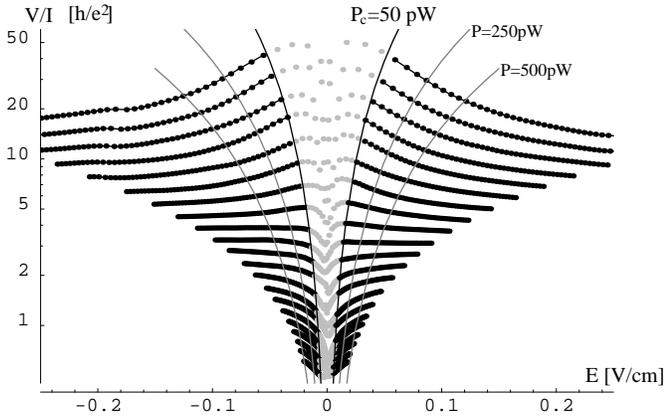}}
  \caption{Nonlinear resistivities $V/I$ for electron
    densities varying from 0.90 to $1.14*10^{11}$ cm$^{-2}$. Here
    a low power cut-off $P_c=50$pW is introduced, discarding all
    data taken at a total power of 50pW or less dissipated in the
    sample. Other possible low power cuts are indicated in grey.
    Here no contact correction for the anomalies at the center of
    the curves is used.}
  \label{Ratios}
\end{figure}

Qualitatively, as has been reported previously,
\cite{kravchenko96a} two different nonlinear $V(I)$ behaviors can
be seen. At low densities the resistivity has a maximum at zero
electric field and decreases for stronger fields. This is similar
to a normal insulator where a larger electric field can excite
more carriers and hence increase the conductivity. At high
electron densities the opposite behavior is seen; the resistivity
has a minimum at $E=0$ and increases for stronger fields. Between
these behaviors, a linear $V(I)$ is observed with a constant
$\rho$, see Fig.~\ref{Ratios}.
This is similar to the nonlinear region of a superconductor.
However, the nonlinearities seen here are anomalously strong,
they occur at very small fields, and they are unexplained by a
clear microscopic model.

Because the \IV curves are nonlinear, there is a difference
between the normal resistivity given by the ratio $V/I$, and the
differential resistivity, $\frac{dV}{dI}$. We have analyzed both,
and they have qualitatively similar behaviors. The differential
resistivity has the advantage of being insensitive to
experimental voltage and current offsets that must be considered
when calculating ratios. The $\frac{dV}{dI}$ is also more
sensitive both to structure in the \IV and to noise. We generate
differential resistivities by doing a n-point differentiating
convolution of the \IV data;
\begin{equation}
\label{convolution}
  \rho_i=\sum_{j=1}^{n} c_j ( V_{i+j} -V_{i-j}) \ ,
\end{equation}
where the data points, $\{V_i\}$, are made equidistant in current
excitation by spline interpolation. Choosing different
Super-Lancos differentiation filters, $\{c_n\}$, allows us to
optimize the data smoothing to reduce noise while retaining the
signal. We then propage $\sigma_V$ in the original data through
the convolution to obtain the variance in the resistance,
$\sigma_\rho$.

\section{Scaling model}
The question at issue is whether or not a quantum ($T=0$) phase
transition between two different conductivity states takes place
at $B=0$ in these systems. As mentioned in the introduction this
is still highly debated in spite of large experimental and
theoretical efforts. Our aim here is to develop a more rigorous
and objective way to determine the consistency of experimental
data with such a phase transition.

A quantum phase transition occurs at zero temperature whereas all
conductivity measurements are taken at both finite bath
temperature and nonzero excitation voltage to drive a current
through the sample. Theoretical arguments suggest that an
underlying quantum phase transition should give rise to {\em
  scaling} at finite temperatures and excitation $E$-fields,
\cite{sondhi97} i.e. that the resistivity of the sample should
have a specific temperature and $E$-field dependence;
\begin{equation}
\label{ETscaling}
  \rho(n_s,T,E)=\rho(x,y) = f \left( \frac{\delta}{E^{1/\beta}},
    \frac{\delta}{T^{1/\nu z}} \right) \ , 
\end{equation}
where $\delta = \frac{n_s - n_c}{n_c}$ is the relative deviation
of the electron density from a critical density $n_c$, and $\nu$,
$z$ and $\beta$ are critical exponents associated with the phase
transition. Assuming the simplest $E$-field heating mechanism,
$\beta$ would be related to the temperature exponents as $\beta=
\nu (z+1)$. Here $x=\delta/E^{1/\beta}$ and $y=\delta/T^{1/\nu
  z}$ are called the scaled variables.

In cases where either the $E$-field or the temperature has been
judged negligible compared to the other, data has been examined
for one-parameter scaling, i.e. whether or not it collapses to a
function of only one scaled variable. At the core of all
arguments for a phase transition lies the fact that one in many
cases can find a critical density $n_c$ and exponent, $\nu z$ or
$\beta$, so that data falls onto a single curve. This is usually
done by trying out different critical densities and exponents.
The best fit is then determined by eye as the parameter choice
that gives the best data collapse.  Because the question of a
phase transition in our system still is controversial, we
developed an objective method to determine the compatibility of
the data with scaling.

Our goal is now to give a {\em quantitative} measure of the
consistency of the data with a single curve when we plot the
resistivities taken at the lowest temperature as a function of
the scaled $E$-field, $x=\frac{(n_s - n_c)}{n_c} E^{-1/\beta}$.

We want to assume as little as possible about the form of the
scaling function, $f(x)$, but previous investigations suggest
that $Log[f(x)]$ is a reasonably smooth curve without too much
curvature. We therefore investigate a model where $Log[f(x)]$ is
a polynomial with only a few terms, $f(x)=\exp [a_0 + a_1 x +a_2
x^2 +a_3 x^3 + \dots]$. Because theory does not tell us the form
of $f(x)$, any functional form for $f$ that is general enough is
acceptable, provided it contains a small number of free
parameters compared to the number of data points being fitted.
If we truncate this polynomial at say third order, we obtain a
scaling model for $\rho(n_s,E)$ that contains the six parameters
$n_c, \beta, a_0, a_1, a_2$ and $a_3$;
\begin{eqnarray}
\label{modelrho}
  \rho_{model}(n_s,E)&=&e^{a_0 + a_1 x +a_2 x^2 +a_3 x^3} \nonumber 
 \, \, \, \mbox{where}\\
  x&=&\frac{(n_s - n_c)/n_c}{E^{1/\beta}} \, .
\end{eqnarray}
We can now address the question of the consistency of the data
with scaling by investigating our ability to fit the measured
data to this scaling model. We do this with a nonlinear
regression routine in {\sl Mathematica$^{TM}$} that fits all
parameters at once and hence obtain values for $n_c$ and $\beta$
with statistical error estimates. At the same time we get an
objective measurement of the 'goodness of fit' via the $\chi^2$
of the fit.

In order to have an unbiased fit and get $\chi^2$ in the correct
units we need to keep track of the estimated uncertainty for each
experimental point. In this case we need to consider not only the
error in $\rho$ itself but also the uncertainty in $n_s$ and in
$E$. We find that the uncertainty in density, $\sigma_n$ normally
dominates the uncertainty of points in the $f(x)$-plot. To take
this into account we propagate all errors to a single effective
error in the resistivity, $\sigma_{\rho,\mbox{\small eff}}$:
\begin{eqnarray}
  \sigma_{\rho,\mbox{\small eff}}^2&=&\left[\sigma_\rho \right]^2 + 
  \left[\sigma_n \frac{\partial \rho_{model}}{\partial n_s} \right]^2
  +\left[ \sigma_V \frac{\partial \rho_{model}}{\partial V} \right]^2
  \ .
\end{eqnarray}
We estimate the error in the resistivity $\sigma_\rho$ by
propagating the errors from the probe voltage $\sigma_V$ either
via the convolution (\ref{convolution}) or the ratio $V/I$. When
calculating the total effective error $\sigma_{\rho,\mbox{\small
    eff}}$, we still assume $\sigma_\rho$ and $\sigma_V$ to be
independent. We then weigh the points in the fit with a weight
$1/\sigma_{\rho,\mbox{\small eff}}^2$. Because the weights depend
on the fitted polynomial, we have to start with an unweighted
fit, then propagate the errors and iterate the fit until a
self-consistent solution is found.

\section{Results}
Using either the derivative $dV/dI$ or the ratio $V/I$ as the
resistivity $\rho$, we have used the scaling model described in
the previous section to investigate $E$-field scaling in three
different samples. Here we used a fifth order polynomial
(\ref{modelrho}). We also achieved reasonable results with lower
order polynomials. Because data is taken at a finite temperature,
($\sim 0.2$K) we know that the resistivity is dominated by the
temperature for small enough excitation $E$-fields; hence, we
introduce a low power cut-off to look for scaling in the
$E$-field only.
 
\begin{figure}[hbtp]
  \epsfxsize 9cm \centerline{\epsffile{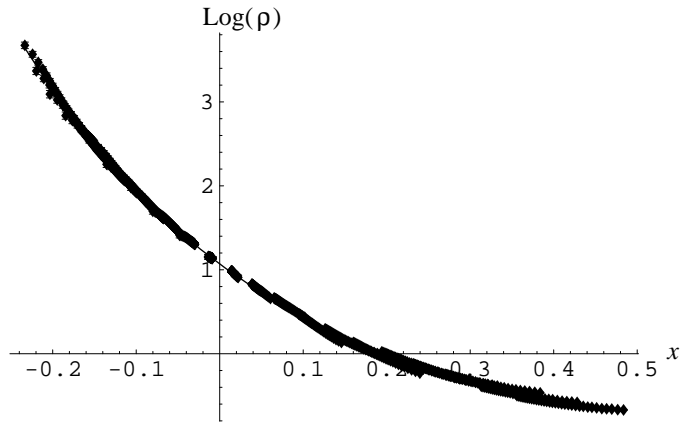}}
  \caption{Scaled data points, $\rho(x)=V/I$, from sample one
    and fitted fifth order model polynomial. The fitted
    polynomial is only visible where it is not covered by data
    points. Here a low power cut-off of 50pW was used and the fit
    has $\chi^2 = 1.86$.}
  \label{scalingfit}
\end{figure}

Using the ratio $V/I$ for the resistivity $\rho$ we obtain a
$\chi^2$ reasonably close to one. Figure \ref{scalingfit} shows
the result from sample one,
using a low power cut-off $P_c=50$~pW.  The critical exponent we
find in this way is comparable to previous
investigations,\cite{kravchenko96a,kravchenko95b,heemskerk98}
even though the exponent and $\chi^2$ do depend somewhat on the
particular cut-offs used.

We have most data from sample one. It also shows the best
behavior; $\chi^2$ is about one and the physically interesting
fit parameters, in particular the exponent $\beta$, are
relatively stable as the low-power cut-off is increased, see
Table \ref{si12ltable}.
\begin{table}[h]
  \begin{tabular}[h]{lllll}
    $P_c$ & $n_c$ & $\beta$ & $a_0$ & $\chi^2$ \\
    \hline
    50  &$1.0142 \pm 0.0006$&$3.83 \pm 0.03$&
    $1.074 \pm 0.009$& 1.86  \\
    100 &$1.0138 \pm 0.0006$&$3.76 \pm 0.03$&
    $1.083 \pm 0.009$& 1.18 \\
    200 &$1.0134 \pm 0.0006$&$3.75 \pm 0.04$&
    $1.094 \pm 0.009$& 0.70 \\
    400 &$1.0136 \pm 0.0009$&$3.81 \pm 0.05$&
    $1.093 \pm 0.013$& 0.45 \\
    500 &$1.0143 \pm 0.0010$&$3.89 \pm 0.06$&
    $1.084 \pm 0.014$& 0.38 \\
  \end{tabular}
  \caption{Results for sample one. Listed are the
    low-power cut-off $P_c$ in pW, the critical 
    density $n_c$ in units of $10^{11}$ cm$^{-2}$, the 
    logarithm $a_0$ of the critical resistivity in 
    $\log h/e^2$ and the average $\chi^2$ for the fit.
    The errors are estimated from the statistical
    deviations of the data from the fitted model only.}
  \label{si12ltable}
\end{table}
The results in the other samples show stronger dependence on the
cut-off. Only for quite large low power cut-offs ($\sim 500$pW)
are the exponents from different samples compatible with each
other, see Table \ref{si13table}
\begin{table}[h]
  \begin{tabular}[h]{lllll}
    $P_c$& $n_c$ & $\beta$ & $a_0$ & $\chi^2$ \\
    \hline
    50  &$0.9717 \pm 0.0004$&$3.14 \pm 0.02$&$2.164 \pm 0.006$& 0.94 \\
    100 &$0.9751 \pm 0.0004$&$3.25 \pm 0.03$&$2.117 \pm 0.006$& 0.63 \\
    200 &$0.9780 \pm 0.0006$&$3.36 \pm 0.04$&$2.078 \pm 0.008$& 0.46 \\
    400 &$0.9818 \pm 0.0010$&$3.53 \pm 0.08$&$2.028 \pm 0.013$& 0.37 \\
    500 &$0.9826 \pm 0.0014$&$3.69 \pm 0.13$&$2.018 \pm 0.017$& 0.36
  \end{tabular}
  \caption{Results for sample two.}
  \label{si13table}
\end{table}
for results from sample two. We note that at an excitation power
of 1~nW temperature increases in the sample were observed via the
temperature monitoring RuO$_x$ resistor. Thus, data at these high
excitations may correspond to the hydrodynamic
regime\cite{chow96,chow97} discussed in this context
earlier.\cite{kravchenko96a}

Sample one shows rather good consistency with the scaling model.
Here all the fitted parameters are reasonably independent of the
low power cut-off, $P_c$. The number of points in a fit varies
from 864 for the smallest cut-off to 467 for the largest. Actual
variations in the fitted parameters are consistent with the
uncertainty estimates based on the statistical deviations from
the model as the low power cut-off is changed. For example, the
critical density $n_c$ is virtually unchanged for all cut-offs.
This is also reflected in very small statistical uncertainties
for $n_c$ in each fit. The exponent $\beta$ (=$\nu~(z+1)$) is
consistent with $\beta=3.8 \pm 0.05$, and the critical
resistivity $e^{a_0}$ is approximately $2.9 h/e^2$. The mean
square deviation per point, $\chi^2$, however, does vary
substantially with the cut-off, decreasing as the cut-off is
increased. This could be understood as an effect of the finite
temperature the data is taken at; at low powers the temperature
is not negligible and the model (\ref{modelrho}) is too crude to
give a good fit.  It is also possible that there are two regimes
of nonlinearity as mentioned above.  The fact that $\chi^2$ is
less than one for the best fits may indicate that we have
overestimated the statistical uncertainties in the data.

Sample two shows a larger dependence on the low power cut-off,
and only at large cut-offs ($\sim 400$pW) is the exponent $\beta$
compatible with the results in sample one. Clearly the exponent
varies much more than the statistical error estimates in each of
the fits with lower cut-offs. Again $\chi^2$ varies in the same
way as in sample one. At the largest cut-offs, however, both
$\beta$ and $\chi^2$ are close to the values found in sample one.
The critical resistivity is, however, clearly not the same as in
sample one, $e^{a_0} \approx 7.5 h/e^2$.

Sample three shows a behavior similar to sample one (exponent
$\beta \approx 3.7$ and $e^{a_0} \approx 3.9$) but with much
poorer statistics.

\begin{figure}[hbtp]
  \epsfxsize 9.5cm \centerline{\epsffile{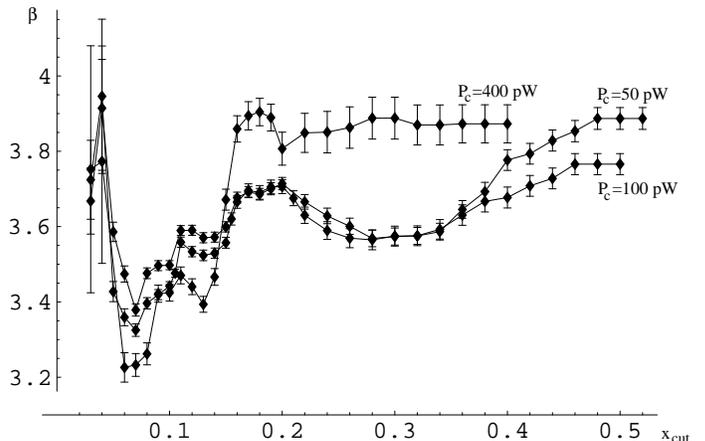}}
  \caption{Fitted exponent $\beta$ with statistical error
    estimates versus $x_{cut}$ for sample one. Here each point
    results from one fit using a cut-off $|x| \le
    x_{\mbox{\footnotesize cut}}$ in the scaled variable $x$.
    Points with with the same low power cut-off $P_c$ are joined
    as indicated in the figure.}
  \label{beta.vs.xcut}
\end{figure}

In order to gain a further understanding of the data, and our
fitting procedure, we have also studied the effect of introducing
a cut-off in the scaled variable $x$. By keeping data where $|x|
\le x_{\mbox{\footnotesize cut}}$ we effectively cut away points
that are taken far from the critical density.  Note that the
criteria for a specific density to be ``near'' or ``far'' from
the critical density is determined also by the excitation field;
a small field corresponds to a more narrow range of densities.
Since $x$ depends also on the critical density and exponent given
by the fit, this cut has to be made self-consistently. Ideally
one would expect fitted parameters to be independent of the
actual value of $x_{\mbox{\footnotesize cut}}$, provided
$x_{\mbox{\footnotesize cut}}$ is not too large.
Figure~\ref{beta.vs.xcut}
shows how the fitted exponent $\beta$ varies with
$x_{\mbox{\footnotesize cut}}$ in sample one. Each point in the
figure corresponds to a complete fit using both the $|x| \le
x_{\mbox{\footnotesize cut}}$ cut-off and a low power cut-off
($P_c$). The statistical uncertainty in each fit, shown as
errorbars on the points, are clearly smaller than the actual
variation in $\beta$. This behavior may be an indication that the
idea of multiple regimes is appropriate, but further
investigations are clearly necessary to settle the question. The
large uncertainties in the points with very small
$x_{\mbox{\footnotesize cut}}$ reflect the fact that there are no
longer enough data left for a good fit.

We have also investigated the derivative $dV/dI$ for scaling. In
this case we find similar but more varied results with generally
higher $\chi^2$. The $\chi^2$'s determined for these fits also
depend on the specific Super-Lancos filter used
(\ref{convolution}).  The derivative scaling also requires the
introduction of cut-offs for high power data. This was done by
discarding data outside three times the FWHM (Full Width Half
Maximum) of the central structure.  Results depend strongly on
this FWHM-cut. The results, presented above, based on ratios show
very little dependence on such high power cut-offs and we have
used all data consistent with other cut-offs.

\section{Conclusions}
We have developed and used a general scaling model to fit all
parameters in $E$-field scaling at once. We find some results
that are consistent with scaling, $\chi^2$ of about one, but also
surprisingly large variations in the fitted parameters when
different physically appropriate cut-offs are introduced.

It would be desirable to generalize and use this method for a
full two-parameter scaling test, treating the $E$-field and
temperature at equal footing. Any one-parameter scaling suffers
from the fact that the other (unscaled) parameter is not zero. As
we have discussed above, any $E$-field scaling will always suffer
from the finite temperature of the surrounding bath. In the same
way, temperature scaling is sensitive to the nonzero excitation
field used to measure resistivities, and heating of the 2DES will
always be a problem for low enough temperatures. Both these
problems are reduced in a two-parameter scaling test where the
scaling model (\ref{ETscaling}) is supposedly valid for any $E$
and $T$ such that $x$ and $y$ are close enough to the transition.

\section{Acknowledgements}
The data used in this study was amassed at the University of
Oklahoma between 1993 and 1996 under the auspices of the
Laboratory for Electronic Properties of Materials, which was
supported by NSF grants EPSCoR-92-OU-NSF and
EPSCoR-95-OUMATERIALS, and The Department of Physics and
Astronomy, which was supported by NSF grants DMR-89-22222 and
DMR-96-264699.  Whitney Mason, Sheena Q.~Murphy, and Sergey
V.~Kravchenko were active collaborators with JEF in collecting
this data.  We wish to thank Kieran Mullen, Bruce Mason, Anders
Karlhede and Christian Walck for many useful discussions.  We
also wish to thank Shankar Das Sarma and the other participants
of the 1998 Program on Disorder and Interactions in Quantum Hall
and MesoScopic Systems at the University of California, Santa
Barbara\cite{itp98} for suggesting this project and providing
much of its initial impetus.


\begin{thebibliography}{10}
  
\bibitem{kravchenko94d} S.~V. Kravchenko, G.~V. Kravchenko, J.~E.
  Furneaux, V.~M. Pudalov, and M.~D'Iorio.  \newblock {\em Phys.\ 
    Rev.~B}, 50:15197, 1994.
  
\bibitem{abrahams79} E.~Abrahams, P.~W. Anderson, D.~C.
  Licciardello, and T.~V. Ramakrishnan.  \newblock {\em Phys.\ 
    Rev.\ Lett.}, 42:673, 1979.
  
\bibitem{kravchenko95b} S.~V. Kravchenko, Whitney Mason, G.~E.
  Bowker, J.~E. Furneaux, V.~M. Pudalov, and M.~D'Iorio.
  \newblock {\em Phys.\ Rev.~B}, 51:7038, 1995.
  
\bibitem{kravchenko96a} S.~V. Kravchenko, D.~Simonian, M.~P.
  Sarachik, Whitney Mason, and J.~E.  Furneaux.  \newblock {\em
    Phys.\ Rev.\ Lett.}, 77:4938, 1996.
  
\bibitem{popovic97} D.~Popovi\'{c}, A.~B. Fowler, and
  S.~Washburn.  \newblock {\em Phys.\ Rev.\ Lett.}, 79:1543,
  1997.
  
\bibitem{heemskerk98} R.~Heemskerk and T.~M. Klapwijk.  \newblock
  {\em Phys.\ Rev.~B}, 58:R1754, 1998.
  
\bibitem{lam97} J.~Lam, M.~D'Iorio, D.~Brown, and H.~Lafontaine.
  \newblock {\em Phys.\ Rev.~B}, 56:R12741, 1997.
  
\bibitem{coleridge97} P.~T. Coleridge, R.~L. Williams, Y.~Feng,
  and P.~Zawadzki.  \newblock {\em Phys.\ Rev.~B}, 56:R12764,
  1997.
  
\bibitem{simmons98} M.~Y. Simmons, A.~R. Hamilton, M.~Pepper,
  E.~H. Linfield, P.D. Rose, D.~A.  Ritchie, A.~K. Savchenko, and
  T.~G. Griffiths.  \newblock {\em Phys.\ Rev.\ Lett.}, 80:1292,
  1998.
  
\bibitem{papadakis98} S.~J. Papadakis and M.~Shayegan.  \newblock
  {\em Phys.\ Rev.~B}, 57:R15068, 1998.
  
\bibitem{ribeiro99} E.~Ribeiro, R.~D Jäggi, T.~Heinzel,
  K.~Ensslin, G.~Medeiros-Ribeiro, and P.~M.  Petroff.  \newblock
  {\em Phys.\ Rev.\ Lett.}, 82:996, 1999.
  
\bibitem{hanein98b} Y.~Hanein, U.~Meirav, D.~Shahar, C.~C. Li,
  D.~C. Tsui, and H. Shtrikman.  \newblock {\em Phys.\ Rev.\ 
    Lett.}, 80:1288, 1998.
  
\bibitem{hanein98a} Y.~Hanein, D.~Shahar, J.~Yoon, C.~C. Li,
  D.~C. Tsui, and H. Shtrikman.  \newblock {\em Phys.\ Rev.~B},
  58:R13338, 1998.
  
\bibitem{QHE} R.~E. Prange and S.~M. Girvin, editors.  \newblock
  Graduate Texts in Contemporary Physics.  Springer-Verlag, New
  York, 1990.
  
\bibitem{QUANTLIQ} S.~Das~Sarma and A.~Pinczuk, editors.
  \newblock Wiley, New York, 1995.
  
\bibitem{sondhi97} S.~L. Sondhi, S.~M. Girvin, J.~P. Carini, and
  D.~Shahar.  \newblock {\em Rev.\ Mod.\ Phys.}, 69:315, 1997.
  
\bibitem{dob97} V.~Dobrosavljevi\'{c}, E.~Abrahams, E.~Miranda,
  and S.~Chakravarty.  \newblock {\em Phys.\ Rev.\ Lett.},
  79:455, 1997.
  
\bibitem{he98} S.~He and X.~C. Xie.  \newblock {\em Phys.\ Rev.\ 
    Lett.}, 80:3324, 1998.
  
\bibitem{castellani98} C.~Castellani, C.~Di~Castro, and P.~A.
  Lee.  \newblock {\em Phys.\ Rev.~B}, 57:R9381, 1998.
  
\bibitem{phillips98} P. Phillips, V. Wan, I. Martin, S. Knysh,
  and D. Dalidovich.  \newblock {\em Nature}, 395:253, 1998.
  
\bibitem{belitz98} D. Belitz and T.~R. Kirkpatrick.  \newblock
  {\em Phys.\ Rev.~B}, 58:8214, 1998.
  
\bibitem{si98} Q.~Si and C.~M. Varma.  \newblock {\em Phys.\ 
    Rev.\ Lett.}, 81:4951, 1998.
  
\bibitem{altshuler99} B.~L. Altshuler and D.~L. Maslov.
  \newblock {\em Phys.\ Rev.\ Lett.}, 82:145, 1999.
  
\bibitem{kravchenko99} S.~V. Kravchenko, M.~P. Sarachik, and D.
  Simonian \newblock {\em Phys.\ Rev.\ Lett.}, 83:2091 1999.
  
\bibitem{altshuler99b} B.~L. Altshuler and D.~L. Maslov.
  \newblock {\em Phys.\ Rev.\ Lett.}, 83:2091, 1999.
  
\bibitem{klapwijk99} T.~M. Klapwijk and S.~Das~Sarma.  \newblock
  {\em Sol.\ St.\ Commun.}, 110(10):581, 1999.
  
\bibitem{itp98} \newblock The Institute for Theoretical Physics,
  UC, Santa Barbara, had a program on Disorder and Interactions
  in Quantum Hall and Mesoscopic Systems in July through December
  1998 that is available at http://www.itp.ucsb.edu/online/.
  
\bibitem{shahar98} D.~Shahar, M.~Hilke, C.~C. Li, D.~C. Tsui,
  S.~L. Sondi, J.~E. Cunningham, and M.~Razeghi.  \newblock {\em
    Sol.\ St.\ Commun.}, 107(1):19, 1998.
  
\bibitem{mason96phd} W.~E. Mason.  \newblock PhD thesis,
  University of Oklahoma, 1996.
  
\bibitem{kruithof91} G.~H. Kruithof and T.~M. Klapwijk.
  \newblock Plenum Press, New York, 1991.
  
\bibitem {chow96}, E.~Chow, H.~P. Wei, S.~M. Girvin, and
  M.~Shayegan, \newblock {\em Phys.\ Rev.\ Lett.}, 77:1143, 1996.
  
\bibitem{chow97} E. Chow, H.~P. Wei, S.~M. Girvin, W. Jan, and
  J.~E. Cunningham.  \newblock {\em Phys.\ Rev.~B}, 56:1676,
  1997.

\end{thebibliography}
\end{document}